\documentclass{PoS}

\usepackage{epsfig,multicol}
\usepackage{amsmath}
\usepackage{amssymb}

\newcommand{\bel}{\begin{equation}\label}

\newcommand {\beq}{\begin{equation}}
\newcommand {\eeq}{\end{equation}}
\newcommand {\beqa}{\begin{eqnarray}}
\newcommand {\eeqa}{\end{eqnarray}}
\newcommand {\bc}{\begin{center}}
\newcommand {\ec}{\end{center}}
\newcommand {\tr}{{\rm tr\,}}

\newcommand {\ee}{\mbox{e}}

\newcommand {\vev}  [1]{\ensuremath{\langle #1 \rangle}}
\newcommand {\rf}   [1]{(\ref{#1})}
\newcommand {\eexp} [1]{\ensuremath{\mbox{e}^{#1}}}
\newcommand {\tl}      {{\tilde\lambda}}

\def\vs5{\vspace*{5mm}}
\def\vs1{\vspace*{1cm}}
\def\vs2{\vspace*{2cm}}
\def\hs5{\vspace*{5mm}}
\def\hs1{\hspace*{1cm}}
\def\hs2{\hspace*{2cm}}
\def\vs50{\vspace*{50mm}}
\def\vs20{\vspace*{20mm}}

\def\tr{\hbox{tr}}

\newcommand\ptp  [3]{{{\it Prog.\ Theor.\ Phys.\ }{\bf #1} (#2) #3}}

\title{A Study of the Complex Action Problem in a Simple Model for
  Dynamical Compactification in Superstring Theory Using the
  Factorization Method}

\ShortTitle{A study of the complex action problem}

\author{\speaker{Konstantinos N. Anagnostopoulos}
        \thanks{The work of K.N.A.\ was partially funded 
by the National Technical University of
Athens through the Basic Research Support Programme 2008.
The work of J.N.\ is supported in part by Grant-in-Aid 
for Scientific Research 
(No.\ 19340066 and 20540286)
from Japan Society for the Promotion of Science.}\\
        Physics Department, 
        National Technical University of Athens, 
        Zografou Campus, 157--80 Zografou, Greece\\
        E-mail: \email{konstant@mail.ntua.gr}}

\author{Takehiro Azuma\\
        Institute for Fundamental Sciences, Setsunan University,
        17-8 Ikeda Nakamachi, Neyagawa, Osaka 572-8508, Japan\\
        E-mail: \email{azuma@mpg.setsunan.ac.jp}}

\author{Jun Nishimura\\
        High Energy Accelerator Research Organization (KEK) and
        Graduate University for Advanced Studies (SOKENDAI),  
        1-1 Oho, Tsukuba 305-0801, Japan\\
        E-mail: \email{jnishi@post.kek.jp}}

\abstract{The IIB matrix model proposes a mechanism for dynamically
  generating four dimensional space--time in string theory by
  spontaneous breaking of the ten dimensional rotational symmetry
  $\textrm{SO}(10)$. Calculations using the Gaussian expansion method
  (GEM) lend support to this conjecture. We study a simple
  $\textrm{SO}(4)$ invariant matrix model using Monte Carlo
  simulations and we confirm that its rotational symmetry breaks down,
  showing that lower dimensional configurations dominate the path
  integral.  The model has a strong complex action problem and the
  calculations were made possible by the use of the factorization
  method on the density of states $\rho_n(x)$ of properly normalized
  eigenvalues $\tl_n$ of the space--time moment of inertia tensor. We
  study scaling properties of the factorized terms of $\rho_n(x)$ and
  we find them in agreement with simple scaling arguments.  These can
  be used in the finite size scaling extrapolation and in the study of
  the region of configuration space obscured by the large fluctuations
  of the phase. The computed values of $\tl_n$ are in reasonable
  agreement with GEM calculations and a numerical method for comparing
  the free energy of the corresponding ansatze is proposed and
  tested.}

\FullConference{The XXVIII International Symposium on Lattice Field
                Theory, Lattice2010\\ 
		June 14-19, 2010\\
		Villasimius, Italy}

\begin{document}
\section{Introduction}

Matrix models have been studied intensively in the past few years in
the context of non-perturbative formulations of string theory and in
the study of gauge/gravity duality. By dimensionally reducing $D=10$
dimensional $\textrm{U}(N)$ supersymmetric (SUSY) Yang--Mills theories
to zero dimensions, one obtains the IIB Matrix Model
\cite{Ishibashi:1996xs} (IKKT model) which has been proposed
as a non-perturbative definition of IIB superstring theory.  In this
model space--time is represented by the distribution of eigenvalues of
the bosonic matrices, a feature that raises the
possibility of {\it dynamical} compactification of the extra
dimensions by Spontaneous Symmetry Breaking (SSB) of the
$\textrm{SO}(D)$ rotational symmetry of the model. Such a scenario is
plausible as calculations using the Gaussian Expansion Method (GEM)
indicate \cite{Nishimura:2001sx}.

Monte Carlo simulations of matrix models
\cite{Krauth:1998xh,Hanada:2007ti} could play an important role in
understanding string theories in a similar fashion that lattice QCD
has contributed to the understanding of the non--perturbative regime
of quantum field theories. Unfortunately such simulations are plagued
by the complex action problem which arises when one simulates the
system after integrating out the fermionic degrees of
freedom. This problem is particularly important 
in the lattice studies of finite density QCD
\cite{Lombardo:2009tf}. The
factorization method has been proposed in \cite{0108041} as a general
method to reduce the complex action problem and eliminate the overlap
problem, see also \cite{Ambjorn:2003rr}.  The basic idea is to control
an {\it appropriately chosen} variable in order to sample regions of
the configuration space which are hard to sample using reweighting and
whose contribution is crucial in the computation of the physical
observables. The study of the scaling properties of the related
density of states allow for useful extrapolations to the physical
results.

We present preliminary results from calculations performed on a
related zero--dimensional matrix model proposed in \cite{0108070}
which realizes the scenario of dynamical compactification of
space--time dimensions \cite{0108070, 0412194}. The model has a very
strong complex action problem and bears strong similarities to the IIB
matrix model, which makes it a useful playground for testing ideas to
apply on the IIB matrix model and more generally on other interesting
physical systems with a complex action problem. We are able to show
that SSB occurs consistently with the predictions in \cite{0108070,
  0412194}. The scaling properties of the density of states are
studied in detail and are found to agree with simple scaling
arguments. This is possible only by sampling heavily suppressed
regions by using the factorization method and it is crucial in the
extrapolations used in order to compute the expectation values of the
SSB order parameters.

\begin{figure}
    \epsfig{file=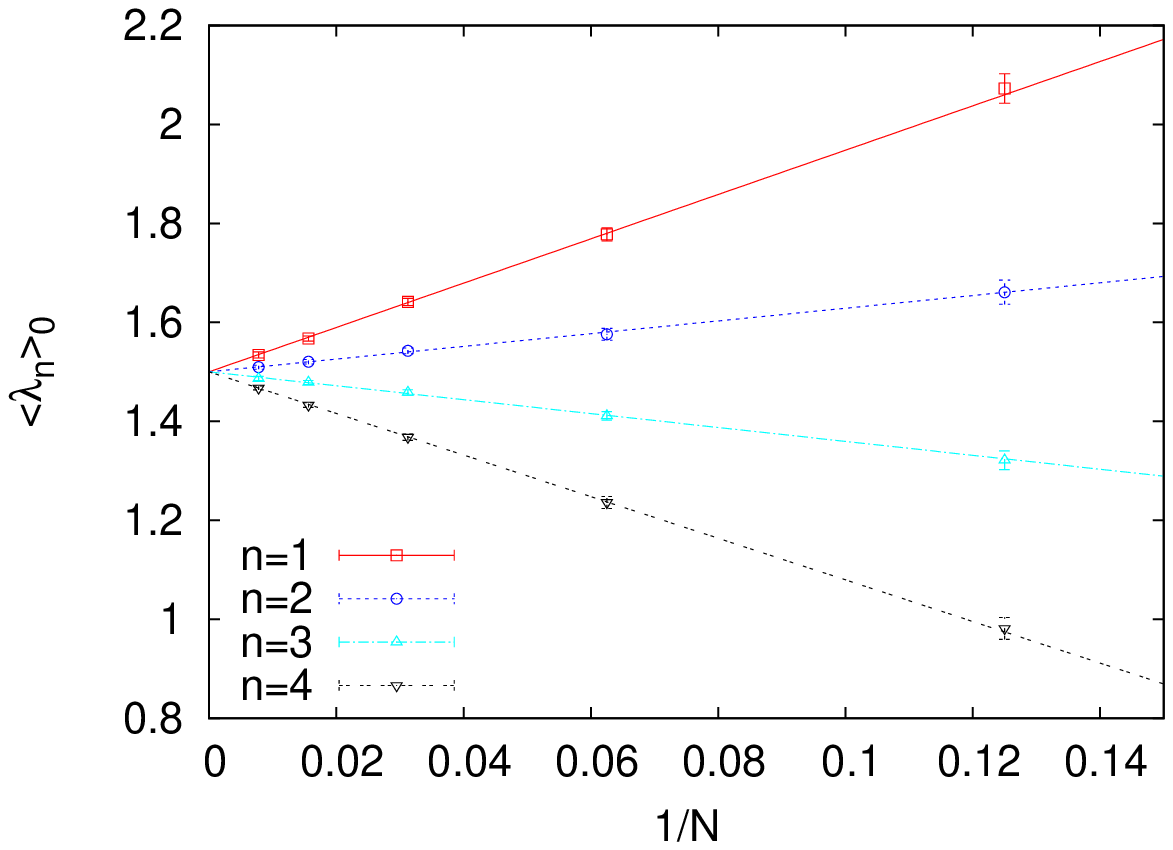,width=7.4cm}
    \epsfig{file=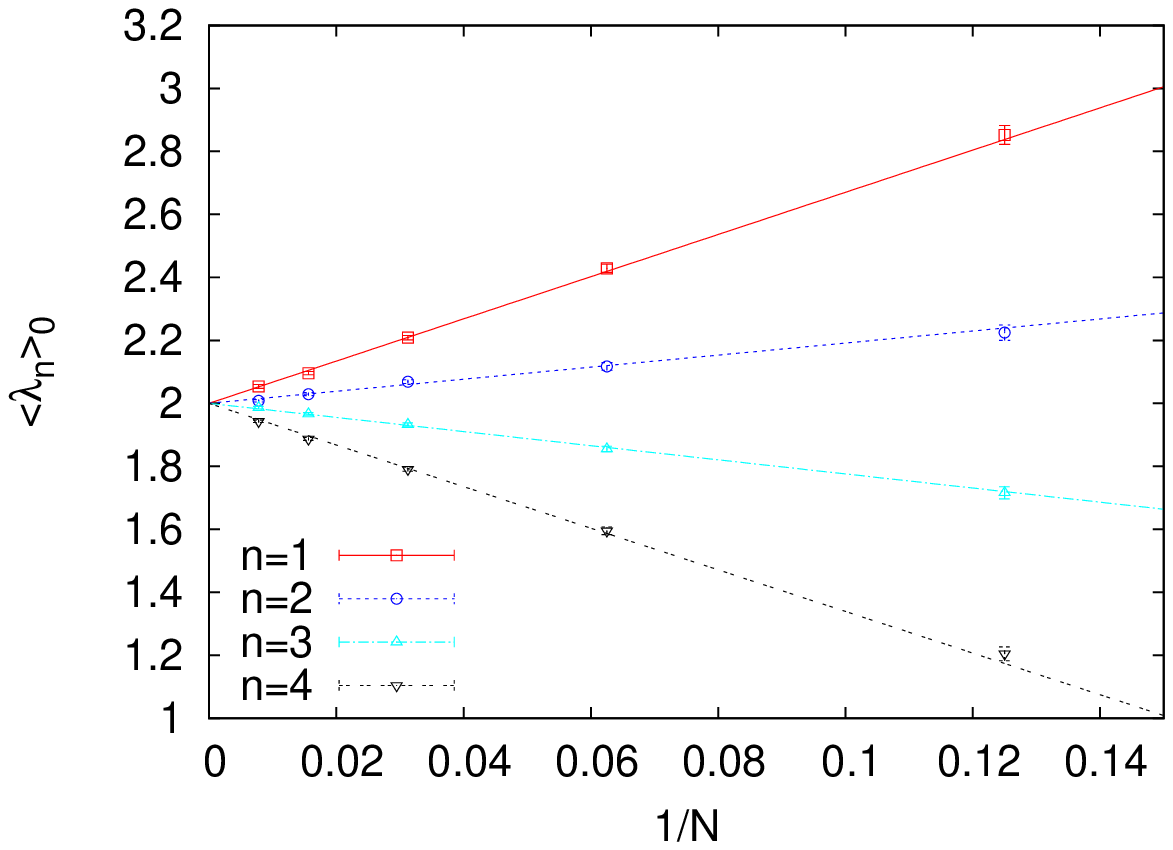,width=7.4cm}
    \caption{The VEV $\langle \lambda_{n} \rangle_{0}$
      ($n=1,2, 3$ and $4$)
in the {\it phase quenched model} $Z_0$ 
are plotted for $r=1$
(left) and $r=2$ (right) against $\displaystyle \frac{1}{N}$. 
The data for each $r$ can be nicely fitted to
straight lines meeting at the same point $(1+r/2)$ at $N=\infty$,
which demonstrates the absence of SSB.}
   \label{no-phase-li}
\end{figure}
\begin{figure}
    \epsfig{file=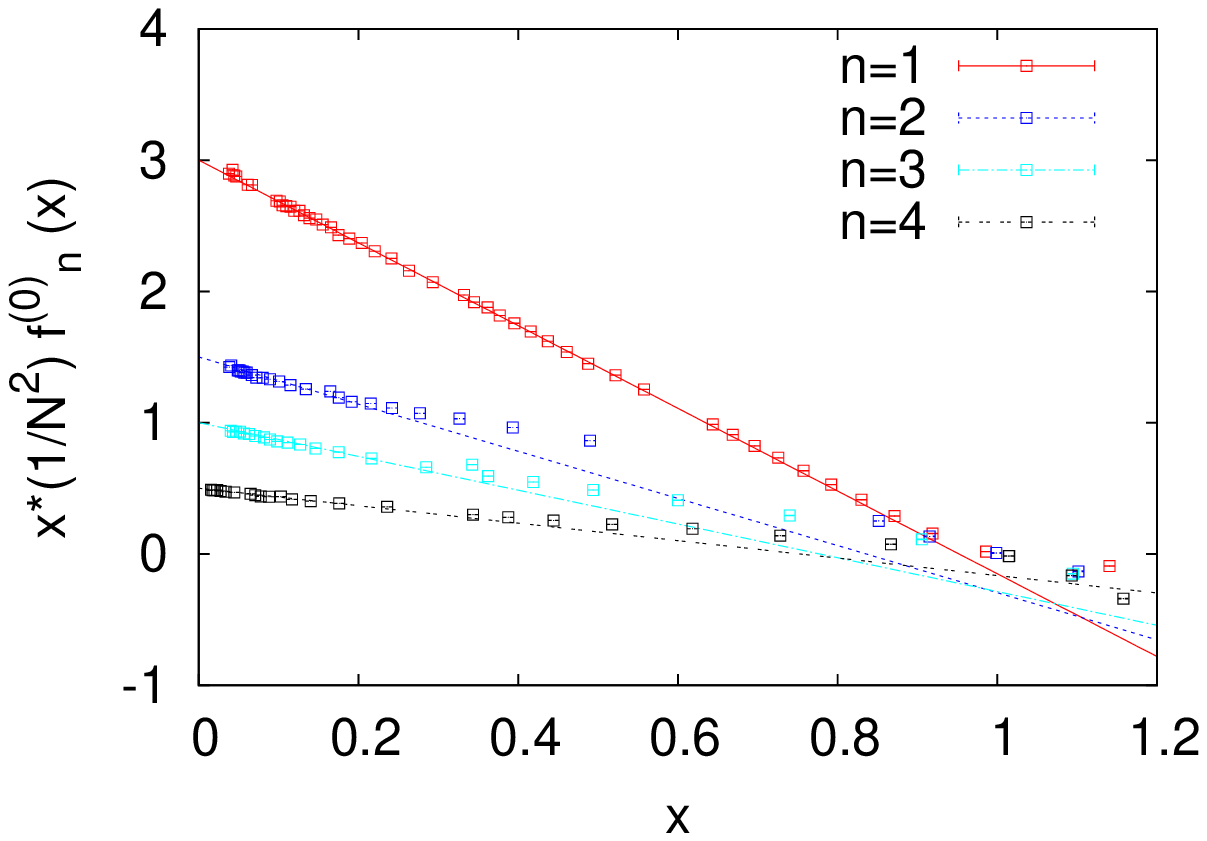,width=7.4cm}
    \epsfig{file=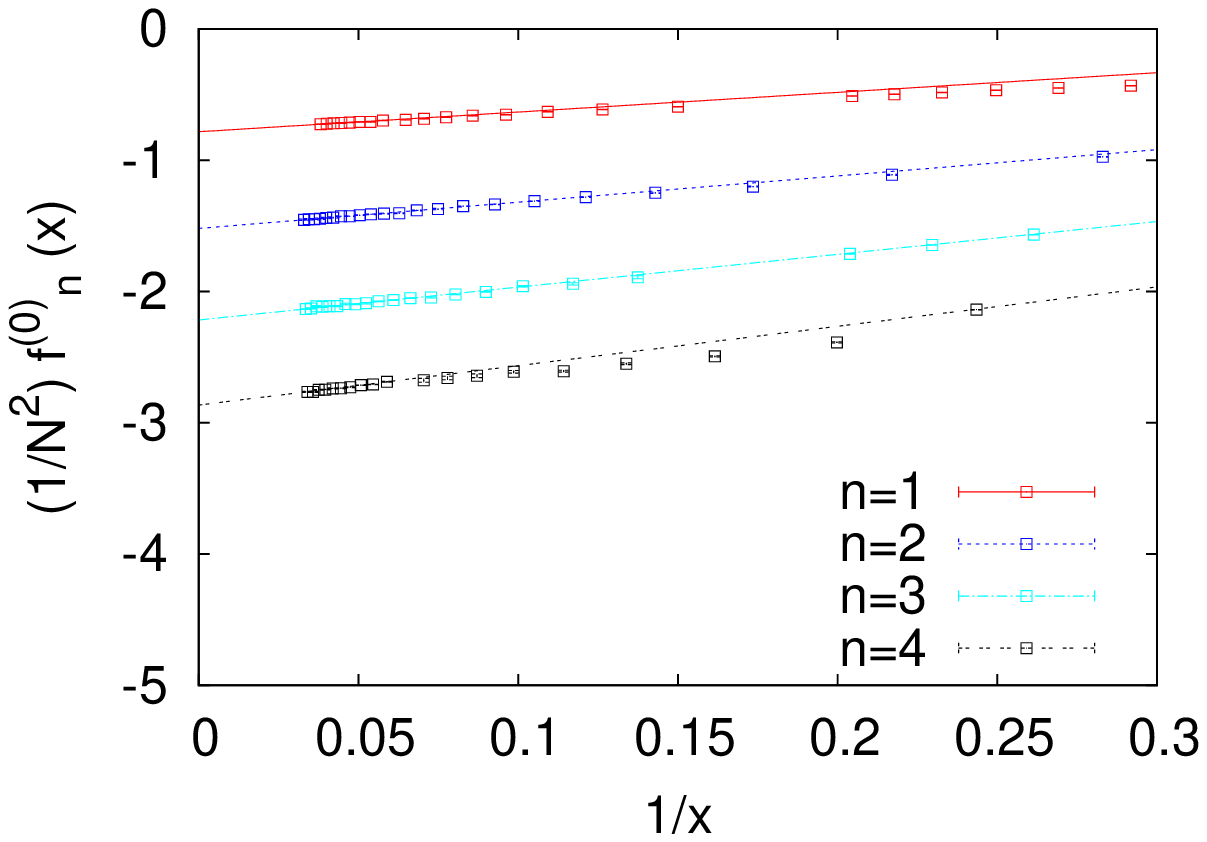,width=7.4cm}
    \caption{The small-$x$ (left) and large-$x$ (right) behavior of
      $\displaystyle{\frac{1}{N^{2}} f^{(0)}_{n}(x)}$ for $r=1$. The straight lines
      are fits to the theoretical behavior \protect\rf{power-law-f}.}
   \label{f-loglog}
\end{figure}
\begin{figure}
    \epsfig{file=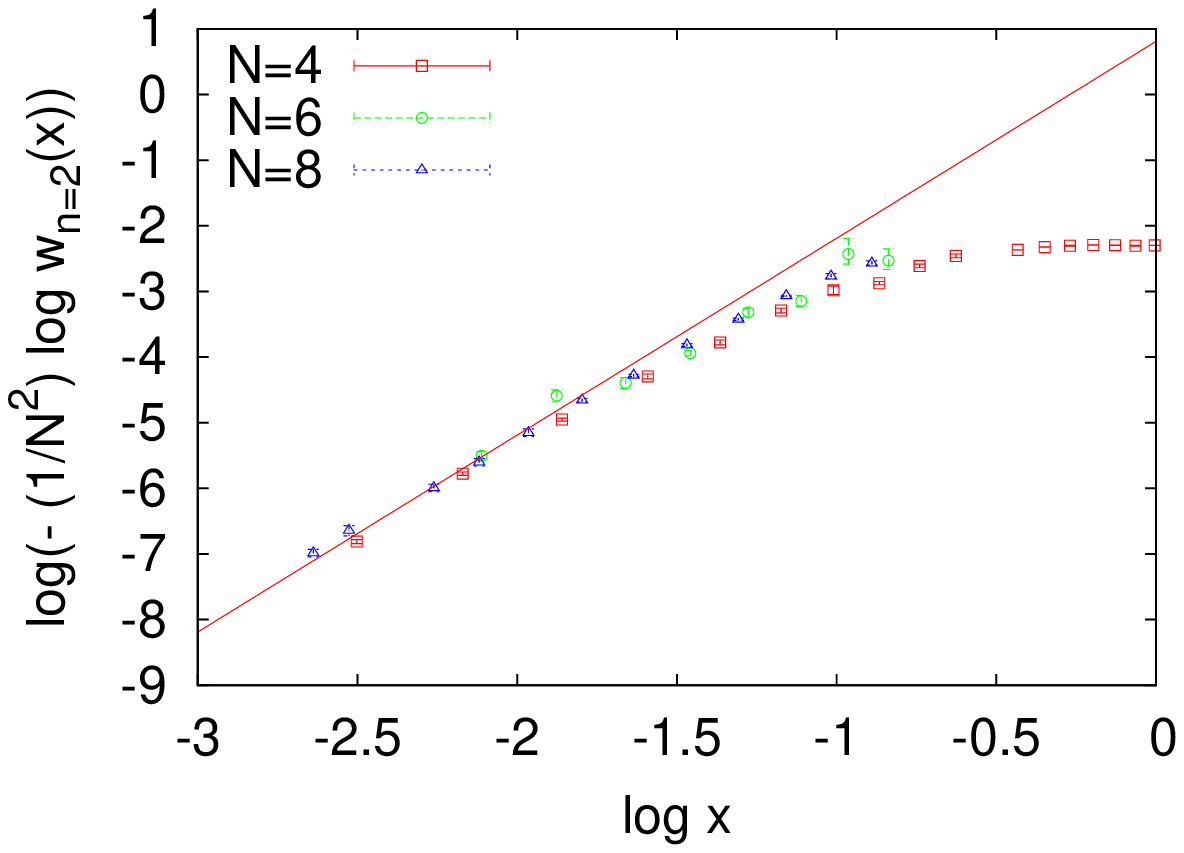,width=7.4cm}
    \epsfig{file=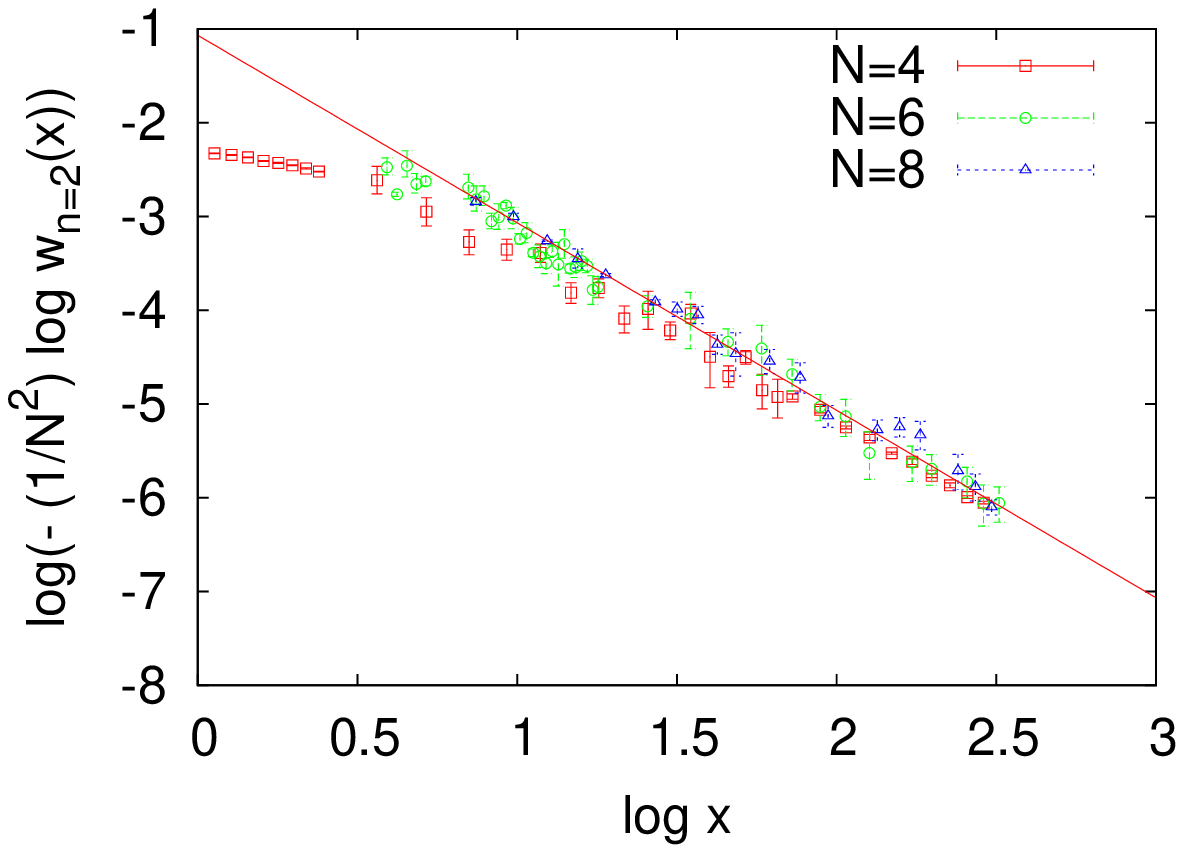,width=7.4cm}
  \caption{The asymptotic behavior (\protect\ref{ii-fit})  for $r=1$ and
    $n=2$.The straight lines are fits to the predicted power-law
behavior using $N=8$ data. The same power law is obeyed also by
smaller $N$ data, 
and a clear trend towards large-$N$ scaling is observed. }  
   \label{res_r1}
\end{figure}

\begin{figure}
    \epsfig{file=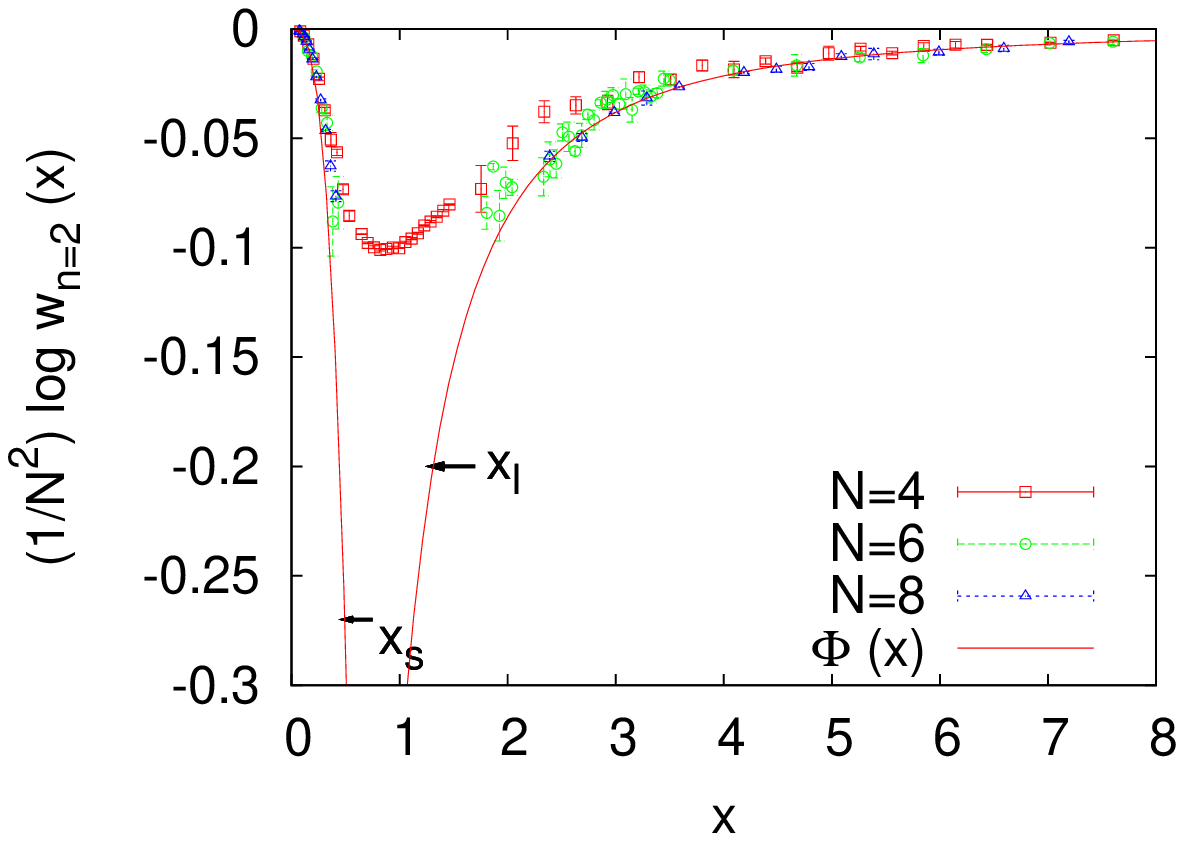,width=7.4cm}
    \epsfig{file=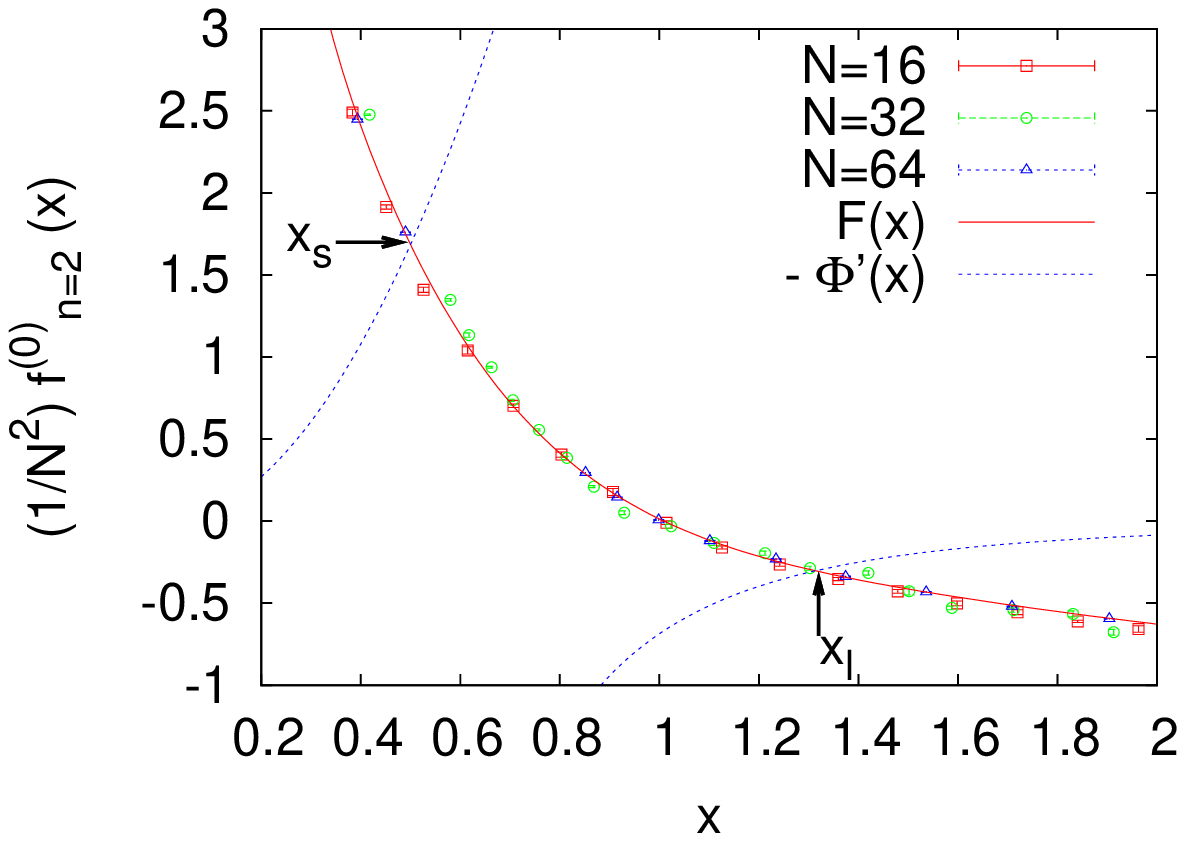,width=7.4cm}
  \caption{(Left) The function $\displaystyle{\frac{1}{N^2} \log w_n (x)}$
is plotted together with the scaling function $\Phi_{n}(x)$ 
extracted from the asymptotic behaviors (\protect\ref{ii-fit}) for $r=1$ and
$n=2$. (Right) The solution to (\protect\ref{eq_xp}) is obtained
by finding the intersections of 
$\displaystyle{\frac{1}{N^{2}} f^{(0)}_{n}(x)}$ and
$\displaystyle{\frac{d}{dx} \Phi_{n}(x)}$ for $r=1$ and
$n=2$. The position of the intersections is indicated by the arrows
with the symbols $x_{\rm s}$  and $x_{\rm l}$
for the regions $x<1$ and $x>1$, respectively.} 
   \label{res_r1-sol} 
\end{figure}

\section{The Model}

Consider the partition function \cite{0108070}
\begin{equation}
\label{total_pf}
  Z = \int dA \, d \psi \, d {\bar \psi} \,
\ee^{-(S_{\rm b} + S_{\rm f})} \qquad\mbox{where}\qquad
  S_{\rm b} = \frac{1}{2} \, N \, \tr (A_{\mu})^{2} \, , \qquad 
  S_{\rm f} = - {\bar \psi}^{f}_{\alpha}\, (\Gamma_{\mu})_{\alpha \beta}\,
  A_{\mu} \psi^{f}_{\beta} \ .
\end{equation}
$A_{\mu}$ ($\mu= 1,\ldots,D$, $D$ even) are $N \times N$ hermitian
matrices, and ${\bar \psi}^{f}_{\alpha}$ and $\psi^{f}_{\alpha}$ are
$N$-dimensional row and column vectors.The actions
$S_{\rm b}$ and $S_{\rm f}$ have an $\textrm{SU}(N)$
symmetry.
The spinor index $\alpha=1,\ldots,p$, where $p$ represents 
the number of components of a $D$-dimensional Weyl spinor,
$p=2^{D/2-1}$, and the flavor index $f=1,\ldots,N_{\rm f}$, where
$N_{\rm f}$ represents the number of flavors. 
The $p \times p$ matrices $\Gamma_\mu$ are $\textrm{SO}(D)$
gamma matrices after the Weyl projection.
Thus the actions (\ref{total_pf}) 
have an $\textrm{SO}(D)$ symmetry,
where the bosonic variables $A_\mu$
transform as vectors and 
the fermionic variables transform as Weyl spinors.
Integrating out the fermions, we obtain
$Z = \int dA \, \eexp{-S_{\rm b}} \, Z_{\rm f}[A]$, 
where 
$Z_{\rm f}[A]= (\det {\cal D})^{N_{\rm f}}$ and
${\cal D} = \Gamma_{\mu} A_{\mu}$ is a $pN \times pN$
matrix. The fermion determinant $\det {\cal D}$ for a single flavor
is complex in general. 
Under parity transformation $A_D\to -A_D$, $A_i\to A_i$ ($i
\ne D)$, the fermion determinant transforms as 
$\det{\cal D}\to (\det{\cal D})^*$. 
This implies that $\det{\cal D}$ is real
for configurations with $A_D=0$ and that 
the phase of the determinant becomes stationary
for configurations with $A_D=A_{D-1}=0$.
We take the large-$N$ limit with $r=N_{\rm f}/N$ fixed, which
corresponds to the Veneziano limit.  Whether the SSB of
$\textrm{SO}(D)$ occurs in that limit is the issue we would like to
address.  For that purpose, we consider the ``moment of inertia
tensor''
$ T_{\mu \nu} = \frac{1}{N} \tr (A_{\mu} A_{\nu})$
and its real positive eigenvalues $\lambda_{n}$ ($n=1,\ldots,D$)
ordered as $\lambda_{1} \geq \lambda_{2} \geq \cdots \geq
\lambda_{D}$.  The vacuum expectation values (VEV) of these
eigenvalues $\langle \lambda_{n} \rangle$ play the role of the order
parameters.  If they turn out to be unequal in the large-$N$ limit, it
signals the SSB of $\textrm{SO}(D)$. Consider the $D=4$ case where we have
$p=2$ and $\Gamma_{i} =  \sigma_{i}$, $i=1,2,3$ and $\Gamma_{4} = i
\sigma_{4}$. The ``phase--quenched model'' is defined by
\beq
  \label{eq:z0}
  Z_0 = \int dA \, e^{-S_{0}[A]} \, , \qquad 
  S_{0}[A] = 
S_{\rm b}[A] - N_{\rm f} \log |\det {\cal D}[A]|\, .
\eeq
It is easy to show that the absence of SSB implies 
\begin{equation}
  \label{eq:pqmlam}
\vev{\lambda_n}_0 =
1+\frac{r}{2} \qquad\mbox{for all}\quad n=1,2,3 , 4 \ ,
\end{equation}
where the VEVs $\vev{\ \cdot \ }_0$ are taken with respect to
$(\ref{eq:z0})$, which is confirmed at infinitesimal $r$
\cite{0108070}, and also at $r=1$ and $r=2$ numerically in this work.
In the full model, GEM calculation up to 9-th order \cite{0412194}
indicate that the true vacuum is only $\textrm{SO}(2)$ invariant and
the $\vev{\lambda_n}$ are not all equal.

In order to simulate  (\ref{total_pf}) we rewrite it as
$ Z = \int dA \, \eexp{-S_{0}[A]} \, \eexp{i \Gamma[A]}$.
This system is very hard to simulate using simple reweighting due to the
complex action and overlap problem which make the
simulations of large systems 
exponentially hard. In this work we use the factorization method
proposed in ref.\ \cite{0108041} where one computes the density of
states of a {\it properly chosen} observable by studying a set of
systems where the observable is constrained to a given fixed
value. The choice of the 
eigenvalues $\lambda_n$ is promising since restricting their values to
be large or small favors configurations with relatively small
fluctuations of the phase $\Gamma$. In this respect it is convenient to define 
${\tilde \lambda}_{n} {=} 
\frac{\lambda_{n}}{\langle \lambda_{n} \rangle_{0}}$ and the
density of states
$\rho_{n}(x) = 
\Big\langle \delta (x-{\tilde \lambda}_{n}) \Big\rangle$
and
$\rho_{n}^{(0)}(x) =
\Big \langle \delta (x- {\tilde \lambda}_{n}) \Big\rangle_{0}$.
Then it is easy to show that 
$ \rho_{n}(x) = \frac{1}{C} \, \rho_{n}^{(0)}(x) \, w_{n}(x)$,
where $C {=}\langle e^{i \Gamma} \rangle_{0} = \langle \cos \Gamma
\rangle_{0}$. It follows that $\langle {\tilde \lambda}_{n} \rangle
= \int_0 ^{\infty}  dx \, x \, \rho_{n}(x)$ and the deviation of its
value from one is a measure of the effect of the phase. The function
$w_{n}(x)$ is defined by 
$w_{n}(x){=} \langle e^{i \Gamma} \rangle_{n,x} = \langle \cos \Gamma
\rangle_{n,x}$, where $\langle \  \cdot  \ \rangle_{n,x}$ 
denotes a VEV with respect to
the partition function
$ Z_{n,x} = \int dA \, \ee^{-S_{0}} \, \delta(x- {\tilde \lambda}_{n} )$.
It turns out that $w_n(x)>0$, which simplifies our analysis 
significantly. Using the saddle point approximation,
the problem of determining $\vev{\tl_n}$ 
can be reduced to that of minimizing 
the ``free energy'' $ {\cal F}_{n}(x) = - \log \rho_{n}(x)$ by solving
the saddle point equation
\begin{equation}
\frac{d}{dx} \log \rho_{n}(x)=
 f^{(0)}_{n} (x)+
 \frac{d}{dx} \log w_n(x) =0\, ,
\label{extrema}
\end{equation}
where $f^{(0)}_{n} (x)=\frac{d}{dx} \log \rho_{n}^{(0)}(x)$.
It is important that the errors due to statistics and finite $N$ do
not propagate exponentially to $\vev{\tl_n}$ as a direct computation
would imply.

The implementation of the above system is obtained by studying
$Z_{n,V} = \int dA \, 
e^{- \{S_{0} + V(\lambda_{n}) \} }$, 
where
$V(z) = \frac{1}{2} \gamma \, (z-\xi)^{2}$ and
$\gamma$ and $\xi$ are real parameters. The parameter $\gamma$
controls the position and width of the peak of $\tl_n$ and it is
chosen large enough so that the results become independent of its
value. In our simulations we used $\gamma$ in the range
$10^3$--$10^7$. Using the fact that $\rho_{n,V}(x)$
${=}$ $\Big\langle \delta (x- {\tilde
  \lambda}_{n}) \Big\rangle_{n,V}$  $\propto$ $\rho_{n}^{(0)}(x) \exp
\Big\{ - V \Big(x \langle \lambda_{n} \rangle_{0}\Big) \Big\}$,
where $\langle \ \cdot \ \rangle_{n,V}$ is a VEV with
 respect to $Z_{n,V}$, the  position of the peak of the distribution
 function $\rho_{n,V}(x)$ 
is given by the solution of
\begin{equation}
f_{n}^{(0)}(x) - \langle \lambda_{n} \rangle_{0} \, 
V'\Big(x \langle \lambda_{n} \rangle_{0}\Big) =0 \ .
\label{eq_xp}
\end{equation}
If we denote the solution by $x_p$, we use the estimators
$x_p=\langle {\tilde \lambda}_{n} \rangle_{n,V}$, 
$ w_{n} (x_{p}) = \langle \cos \Gamma \rangle_{n,V}$ and
$ f_{n}^{(0)} (x_{p})
=  \langle \lambda_{n} \rangle_{0}  \, 
V'\Big(\langle \lambda_{n} \rangle_{n,V} \Big) 
= \gamma \, \langle \lambda_{n} \rangle_{0}  \, 
\Big(\langle \lambda_{n} \rangle_{n,V} - \xi \Big)$.

\section{Results}

\begin{table}
\begin{center}
\begin{tabular}{||c||c|c|c|c||c|c|c|c||}
\hline
\hline
    & \multicolumn{4}{|c||}{$r=1$} &\multicolumn{4}{|c||}{$r=2$}\\
\hline
$n$ & $ x_{\rm s}$  & $ x_{\rm l}$  & 
      $ x_{\mbox{\scriptsize{SO(2)}}}$  & 
      $ x_{\mbox{\scriptsize{SO(3)}}}$  & 
      $ x_{\rm s}$  & $ x_{\rm l}$  &  
      $ x_{\mbox{\scriptsize{SO(2)}}}$  & 
      $ x_{\mbox{\scriptsize{SO(3)}}}$  \\ 
\hline
1&                    &{              2.12}&{    1.4}&               1.2 &                    &{              1.94}&{    1.7}&               1.2 \\
2&               0.49 &{\bf           1.29}&{\bf 1.4}&               1.2 &               0.48 &{\bf           1.36}&{\bf 1.7}&               1.2 \\
3&{\bf           0.67}&\underline{\it 1.13}&{\bf 0.7}&\underline{\it 1.2}&{\bf           0.53}&\underline{\it 1.16}&{\bf 0.5}&\underline{\it 1.2}\\
4&\underline{\it 0.75}&                    &     0.5 &\underline{\it 0.5}&\underline{\it 0.51}&                    &     0.1 &\underline{\it 0.3}\\
\hline
\hline
\end{tabular}  
\end{center}
\caption{\label{tab:lam}The solutions ($x_{\rm s}$, $x_{\rm l}$) 
to eq.\ \protect\rf{extrema} that correspond to the (local)
maxima of $\rho_n(x)$ are shown.
We also add the corresponding VEV $\vev{\tl_n}$ obtained by 
the Gaussian expansion method in \protect\cite{0412194}:
$x_{\mbox{\scriptsize{SO(2)}}}\equiv\vev{\tl_n}$ obtained using the
$\mathrm{SO}(2)$ ansatz and 
$x_{\mbox{\scriptsize{SO(3)}}}\equiv\vev{\tl_n}$ obtained using the
$\mathrm{SO}(3)$ ansatz. Bold/italic numbers for same $r$ and $n$ are
to be compared according to the text.}
\end{table}

First we study the phase quenched model $Z_0$. We simulate the system for
$r=1,2$ and compute the eigenvalues $\vev{\lambda_n}_0$. We find that
$\vev{\lambda_n}_0(N) = 1+r/2 + {\cal O}(1/N)$ as can be seen from
fig.  \ref{no-phase-li}. We conclude that no SSB occurs in the phase
quenched model and verify eq. \rf{eq:pqmlam}.  We calculate
$f^{(0)}_n(x)$ from eq. \rf{eq_xp} by simulating $Z_{n,V}$. Using simple scaling
arguments we find that its asymptotic behavior at
$x\ll 1$ and $x\gg 1$ is
\begin{equation}
 \frac{1}{N^{2}} f^{(0)}_{n}(x) \simeq 
 \left\{ 
  \begin{array}{ll} 
   \left(   \frac{1}{2} (5-n)  + r \delta_{n1}  \right) \frac{1}{x} + a_{n}
& x \ll 1  \ , \\ 
    - \frac{1}{2} n \vev{\lambda_n}_0  + \left( \frac{n}{2} + r  \right) \frac{1}{x}
& x \gg 1 \ .
  \end{array} 
  \right. 
\label{power-law-f}
\end{equation}

Similar arguments lead to the respective asymptotic behavior of $w_n(x)$ 
\begin{equation}
\frac{1}{N^2} \ln w_n(x) \simeq \Phi_n(x) = 
  \left\{ 
  \begin{array}{ll} 
    -c_{n} x^{5-n} 
& (x \ll 1, n=2,3,4)  \ , \\ 
    -d_{n} x^{-(4-n)} 
& (x \gg 1, n=1,2,3)  \ .
  \end{array} 
  \right. 
\label{ii-fit}
\end{equation}
By varying the constants $a_n$, $c_n$ and $d_n$, we fit our data to
eqs.  \rf{power-law-f} and \rf{ii-fit}. We
verify the expected asymptotic behaviors and use the coefficients
$c_n$ and $d_n$ in order to extrapolate $\Phi_n(x)$ to the region in $x$
where we find the solution to the saddle point equation \rf{eq_xp}. In
fig. \ref{f-loglog} we show the scaling   \rf{power-law-f} for $r=1$ 
and in fig. \ref{res_r1} the scaling   \rf{ii-fit} for $r=1$ and
$n=2$. The solution to eq.  \rf{eq_xp} is determined from the
intersection of the curves $\frac{1}{N^2}f^{(0)}_n(x)$ with
$-\Phi_n'(x)$ for each $n$. The results for $r=1$ and $n=2$ are shown in
fig. \ref{res_r1-sol}.

We use the notation $x_{\rm s}$ and $x_{\rm l}$ for the solutions in
the $x<1$ and $x>1$ regions respectively which correspond to the local
maxima of $\rho_n(x)$. For $n=1$ we obtain only $x_l\equiv
\vev{\tl_1}$ and for $n=4$ we obtain only $x_s\equiv \vev{\tl_4}$.
We tabulate the results in table \ref{tab:lam} and we compare them
with those obtained using GEM in \cite{0412194}. We note that
since the dominant configurations near $x_l$ for $n=2$ and $x_s$ for
$n=3$ are typically two dimensional, these are to be compared with the
$\textrm{SO}(2)$ ansatz. Similarly, since the dominant configurations near
$x_l$ for $n=3$ and $x_s$ for $n=4$ are typically three dimensional,
these are to be compared with the $\textrm{SO}(3)$ ansatz.
 
We find that $\vev{\tl_1}>1>\vev{\tl_4}$, a relation that is clearly
going to survive the large--$N$ limit. Therefore we conclude that
$\textrm{SO}(4)$ SSB manifests in the model. In order to determine the group
that $\textrm{SO}(4)$ breaks to, we need to calculate the dominant peak as
$N\to \infty$. We consider the quantity 
$\Delta_{n} {=}  
   \frac{1}{N^{2}} 
\Bigl\{ \log \rho_{n} (x_{l}) - \log \rho_{n}(x_{s}) \Bigr\} 
 =  \Phi_{n} (x_{l}) - \Phi_n(x_{s}) +   \Xi_n \, ,$
where $\Xi_n {=}\int^{x_{l}}_{x_{s}} dx \,
\left\{ \frac{1}{N^2} f^{(0)}_n(x)  \right\}$.
If $\Delta_n>0$ the peak at $x_{\rm l}$ dominates, otherwise $x_{\rm
  s}$. We find $\Delta_2\approx 0.34$ for $r=1$ and $\Delta_2\approx
0.25$ for $r=2$ and we conclude that SSB breaks at least down to
$\textrm{SO}(2)$. Unfortunately $\Delta_3$ turns out to be very close to $0$, so we
are unable to determine if $\textrm{SO}(4)$ breaks to $\textrm{SO}(2)$
as GEM predicts 
or to $\textrm{SO}(3)$.

\section{Conclusions}

We have tested a scenario for dynamical compactification of
space--time by simulating a toy matrix model related to the IIB matrix
model of string theory. We have shown SSB of $\textrm{SO}(4)$ rotational
symmetry consistent with GEM analysis and small $r$
calculations\cite{0108070,0412194}. The phase quenched model has no SSB,
confirming the expectation that the wild fluctuations of the phase of
the fermionic partition function plays a crucial role in the mechanism
of SSB. Large and small length scales are dynamically generated by
these fluctuations which make the calculation of the dominant ones in
the thermodynamic limit a challenging problem. Our results indicate
how to proceed with the study of the IIB matrix model.  Although the
latter case is computationally more demanding, SUSY could make SSB
easier to see.

Calculations were possible because of the use of the factorization
method. By effectively sampling large and small $x$ regions we are
able to exploit the asymptotic behaviors of $\rho^{(0)}_n(x)$ and
$w_n(x)$ in order to extrapolate the results to regions in $x$ and
system size which are inaccessible by direct simulations of the phase
quenched model. It is possible that a remaining overlap problem makes
the results of table \ref{tab:lam} slightly differ from GEM
results. By a generalization of the factorization
method\cite{multifac} this problem can be overcome and achieve also
better quantitative agreement. Then Monte Carlo studies of many
interesting systems hindered by the complex action problem are
hopefully going to be made possible by using the factorization method.


\end{document}